\documentclass[
showpacs,preprintnumbers,amsmath,amssymb]{revtex4}
\usepackage{amsmath}
\usepackage{graphicx}
\usepackage{subfig}
\usepackage{hyperref}
\usepackage{amssymb}
\usepackage[english]{babel}
\usepackage{epsfig}
\usepackage{wasysym}
\usepackage{graphicx}
\usepackage{indentfirst}
\usepackage{amsmath}
\usepackage{amsfonts}
\usepackage{amssymb}
\usepackage{enumerate}

\hypersetup{colorlinks,citecolor=green,filecolor=magenta,linkcolor=red,urlcolor=cyan,pdftex}

\newcommand{\be}{\begin{equation}}
\newcommand{\ee}{\end{equation}}
\newcommand{\bea}{\begin{eqnarray}}
\newcommand{\eea}{\end{eqnarray}}
\newcommand{\beaa}{\begin{eqnarray*}}
\newcommand{\eeaa}{\end{eqnarray*}}

\begin{document}


\title{Holographic dark energy model in unimodular $f(T)$ gravity}

 \author{ A. E. Godonou$^{(a)}$\footnote{e-mail: emilog2004@yahoo.fr}, M. J. S. Houndjo$^{(a, b)}$\footnote{e-mail: sthoundjo@yahoo.fr} and
 J. Tossa$^{(a)}$\footnote{e-mail: joeltossa@gmail.com}}


\affiliation{$^a$ \, Institut de Math\'{e}matiques et de Sciences Physiques, 01 BP 613,  Porto-Novo, B\'{e}nin\\
$^{b}$\, Facult\'e des Sciences et Techniques de Natitingou, BP 72, Natitingou, B\'enin} 

\begin{abstract}
The present work deals with holographic dark energy in the context of unimodular $f(T)$ gravity, which is a modification of teleparallel gravity. We develop the general reconstruction procedure of the $f(T)$ form that can yield the holographic feature of the dark energy. We fit the reconstructed model with the $H(z)$ data and our results
show a perfect agreement with the WMAP9 cosmological observational data, at least for the range $-1.10\leq \omega_V \leq -1.05$. We investigate the consistency of the reconstructed model by studying its stability against  linear  gravitational and matter perturbations, fixing $\omega_V$ to $-1.05$. The model presents stability for both de Sitter and power-law solutions and we conclude that it is a good candidate as alternative viable model for characterizing holographic dark energy.

\end{abstract}
\pacs{98.80.-k, 95.36.+x, 04.50.Kd}

\maketitle 

\section{Introduction}
The holographic principle (HP) has been proposed for the first time, through the investigation of black hole thermodynamic \cite{1dewang,2dewang}, by Gerard't Hooft \cite{3dewang}. According to the HP, all of the information contained in a volume of space can be represented as a hologram, view as a theory locating on the boundary of the space. The most well known successful realization of the HP is the famous AdS/CFT correspondence proposed by Maldacena in 1997 \cite{5dewang}, and now it is widely believed to be a fundamental principle of quantum gravity. As some success fields for the HP, the AdS/QCD correspondence has been proposed to explore the problems of quark-gluon plasma \cite{6dewang} in nuclear physics; the AdS/CMT correspondence  has been proposed to study the superconductivity and super-fluid problems in condensed matter physics \cite{7dewang}; the holographic entanglement entropy coming from the Ad/CFT correspondence has been developed in the theoretical physics \cite{8dewang} and the same correspondence allows to discuss the nature of the Sitter space in inflation \cite{9dewang}. It is then obvious that the HP presents great potential to solve many long issues in various physical fields.\par
It is well known nowadays that the exotic component responsible of the cosmic acceleration is the so-called dark energy. This prescription allows to accommodate the alleged accelerated expansion of the universe in the framework of General Relativity (GR). However this prescription is not the unique way to point out the physical properties of the dark energy. The cosmological constant is the first assumption highly consistent with the cosmological observations, which unfortunately suffers from a fine tuning problem \cite{1depurba,2depurba}. Since this problem of fine tuning of cosmological constant is not yet understood, the possible way to bypass this issue is to look for alternative models for the matter content as scalar field, like quintessence \cite{3depurba}-\cite{3depurbaf}, phantom field \cite{4depurba}, tachyon field \cite{5depurba}-\cite{5depurbaf} or fluid models like Chaplygin gas \cite{6depurba}-\cite{6depurbaf}. Still in the optic to get away with fine tuning of cosmological constant, another technique is modifying GR, and thereby the generalized  GR theory, namely $f(R)$ gravity has been proposed with various potential results ($R$ being the curvature scalar), see \cite{fRi}-\cite{fRf} for some of these works; $f(G)$ gravity where $G$ is the Gauss-Bonnet invariant (see \cite{fGi}-\cite{fGf}); $f(R,\mathcal{T})$ gravity where $\mathcal{T}$ is the trace of the stress tensor \cite{fRTi}-\cite{fRTf}. There is also the generalized version of the Tele-parallel (TT), namely $f(T)$ gravity, $T$ being the torsion scalar, where interesting cosmological results have been obtained \cite{fTi}-\cite{fTf}.\par
In this paper, we focus on holographic dark energy, searching for its correspondent $f(T)$ model in a specific way, say unimodular $f(T)$. Note that unimodular gravity \cite{22debamba}-\cite{bamba} is an interesting gravitational theory which can be considered as a specific  case of GR (or TT). As we have noted above the origin of cosmological constant is not well understood, while according to unimodular point of view it arises the trace-free part of the gravitational field equations once the determinant of the metric tensor is fixed to a number. The unimodular presents as great theoretical advantage the fact that since the trace-free part of the field equations in not related to the vacuum expectation value of any matter field, its value can easily be chosen without being confronted to the cosmological constant problem. Therefore the unimodular can be used to describe both the early and late time cosmic regimes of the universe \cite{24debamba,25debamba}. Our task in this paper is to reconstruct the unimodular $f(T)$ model able to reproduce the holographic dark energy feature in agreement with cosmological $H(z)$ data. Moreover, for more consistency, we study the stability of the reconstructed model against linear perturbation through de Sitter and power-law solutions. Our results present viability of the model for some values of the input parameters.\par
The plan of the work is the following:  In Sec. \ref{sec2} we present the general description of unimodular gravity according to FRW metric, and reconstruct the related holographic dark energy model in the framework of $f(T)$ gravity in Sec. \ref{sec3}. We confront the reconstructed model with the observational $H(z)$ in Sec. \ref{sec4} and study its stability against linear perturbation in Sec. \ref{sec5}. The conclusion and perspective are presented in Sec. \ref{sec6}.

\section{General description of unimodular gravity}\label{sec2}
In this section we address the generalization of the GR gravity formalism within unimodular $f(T)$ gravity formalism. Note that unimodular gravity approach is essentially based on the assumption that the determinant cannot change, i.e, the metric tensor is fixed and  generated by the relation $g_{\mu\nu}\delta g^{\mu\nu}=0$. Throughout this paper we fix the metric such that \cite{bamba}
\begin{eqnarray}\label{cond1} 
\sqrt{-g}=1
\end{eqnarray}
In this paper we focus on FWR metric, as 
\begin{eqnarray}\label{metric1}
ds^2=dt^2-a(t)^2dx^2-a(t)^2dy^2-a(t)^2dz^2
\end{eqnarray}
It is obvious from (\ref{metric1}) that the unimodular constraint, expressed by the Eq. (\ref{cond1}) is not satisfied and then, in order to satisfy this later, we redefine the cosmic time coordinate as follows
\begin{eqnarray}
d\tau = a^3(t)dt,\label{ttau}
\end{eqnarray} 
such a way that the metric (\ref{metric1}) becomes
\begin{eqnarray}\label{metric2}
ds^2= a\left[t(\tau)\right]^{-6}d\tau^2-a(t(\tau))^2\left(dx^2+dy^2+dz^2\right).
\end{eqnarray}
The previous metric clearly satisfies the constraint (\ref{cond1}) and we shall refer to it as the unimodular FRW metric. 
\section{Unimodular $f(T)$ gravity with the account of holographic dark energy}\label{sec3}
In this section we start presenting the general $f(T)$ gravity action, coupled with matter
$L_m$ by \cite{fTi}
\begin{equation}\label{fTaction}
S=\frac{1}{16\pi G}\int d^4x e \left[f(T)+L_m\right]_,
\end{equation}
where $e=det(e^i_{\mu})=\sqrt{-g}$. In what follows, we will assume the units $8\pi G=1$. Here $T$ denotes the torsion scalar and is defined as
\begin{equation}\label{torsionscalar}
T=S^{\:\:\:\mu \nu}_{\rho} T_{\:\:\:\mu \nu}^{\rho},
\end{equation}
where
\begin{eqnarray}
T_{\:\:\:\mu \nu}^{\rho}=e_i^{\rho}(\partial_{\mu}
e^i_{\nu}-\partial_{\nu} e^i_{\mu}),\label{Ttensor}\\
S^{\:\:\:\mu \nu}_{\rho}=\frac{1}{2}(K^{\mu
\nu}_{\:\:\:\:\:\rho}+\delta^{\mu}_{\rho} T^{\theta
\nu}_{\:\:\:\theta}-\delta^{\nu}_{\rho} T^{\theta
\mu}_{\:\:\:\theta}),\label{Storsion}
\end{eqnarray}
and $K^{\mu \nu}_{\:\:\:\:\:\rho}$ is the contorsion tensor defined as
\begin{eqnarray}
K^{\mu \nu}_{\:\:\:\:\:\rho}=-\frac{1}{2}(T^{\mu
\nu}_{\:\:\:\:\:\rho}-T^{\nu \mu}_{\:\:\:\:\:\rho}-T^{\:\:\:\mu
\nu}_{\rho}).\label{Ktorsion}
\end{eqnarray}
By varying the action (\ref{fTaction}) with respect to vierbein $e^i_{\mu}$, one gets the general field equations 
\begin {equation}\label{manuel6}
S^{\;\;\nu\beta}_{\mu}\partial_{\beta}(T)f_{TT}+\left[e^{-1}e^{i}_{\mu}\partial_{\beta}\left(ee^{\;\;\alpha}_{i}S^{\;\;\nu\beta}_{\alpha}\right)+T^{\alpha}_{\;\;\lambda\mu}S^{\;\;\nu\lambda}_{\alpha}\right]f_{T}+\frac{1}{4}\delta^{\nu}_{\mu}f=\frac{1}{2}\mathcal{T}^{\nu}_{\mu}\; .
\end{equation}
Here  $f_T$ and $f_{TT}$ denote the first and second derivatives of $f$ with respect to $T$, while $\mathcal{T}_{\rho\nu}$ is the stress tensor, and we also set $\kappa^2=8\pi G=1$. According to the unimodular-like FRW line element (\ref{metric2}) the temporal and space equations of field read
\begin {eqnarray}
T&=&-6a^6\mathcal{H}^2\,\,\,,\label{manuel8}\\
12a^6\mathcal{H}^2f_T+f&=& 2\rho \,\,\,,\label{manuel9}\\
48a^{12}\mathcal{H}^2\left(3\mathcal{H}^2+\mathcal{H}'\right)f_{TT}-4a^{6}\left(6\mathcal{H}^2+\mathcal{H}'\right)f_{T}-f&=&2p            \,\,,\label{manuel10}
\end{eqnarray}
where $\rho$ and $p$ are the energy density and pressure  of ordinary matter content of the universe respectively. The related Hubble parameter is $\mathcal{H}$ (we should call it unimodular Hubble parameter) and defined as
$\mathcal{H}={a}'/a$, where the ``{\it prime}" denotes the derivative with respect to the $\tau$. \par
As we are dealing with holographic dark energy, one can consider it contribution coming from an algebraic function $g(T)$ such that
$f(T)=T+g(T)$,     \;\;$T$ providing the TT theory related to the ordinary content of the universe. Thus, the equations (\ref{manuel9}-\ref{manuel10}) becomes
\begin{eqnarray}
6a^6\mathcal{H}^2&=&2\rho-12a^6\mathcal{H}^2g_T-g\,\,\label{manuel11}\\
-2a^6\left(9\mathcal{H}^2+2\mathcal{H}'\right)&=&2p-48a^{12}\mathcal{H}^2\left(3\mathcal{H}^2+\mathcal{H}'\right)g_{TT}+
4a^6\left(6\mathcal{H}^2+\mathcal{H}'\right)g_T+g\,\,\label{manuel12}
\end{eqnarray}
As well known in the literature there is a correspondence between the  holographic dark energy and the algebraic $f(T)$ dark energy function. The energy density $\rho_V$ related to the holographic dark energy  can be written as \cite{setareft}-\cite{chinois}
\begin{eqnarray}\label{manuel12}
\rho_V=\frac{3b^2}{R^2_h}\,\,\,.
\end{eqnarray}
Here $b$ is a constant whiled $R_h$ denotes  the future event horizon, expressed in terms of cosmic time et $\tau$, respectively, as 
\begin{eqnarray}\label{manuel13}
R_h=a(t)\int^{\infty}_{t}\frac{dt^{\prime}}{a(t^{\prime})}=a[t(\tau)]\int ^{\infty}_{\tau}\frac{d\tau '}{a^4[t(\tau ')]}\,\,\,,\label{Rhdef}
\end{eqnarray}
which can be simply transformed as
\begin{eqnarray}\label{manuel14}
R_h=a\int^{\infty}_{a}\frac{da}{\mathcal{H}a^5}\,\,\,.
\end{eqnarray}
Making use of the critical energy density $\rho_{cr}=3a^6\mathcal{H}^2$ from (\ref{manuel11}), one may define the dimensionless dark energy as
\begin{eqnarray}\label{manuel15}
\Omega_V=\frac{\rho_V}{\rho_{cr}}=\frac{b^2}{a^6\mathcal{H}^2R^2_h}\,\,\,.\label{Omega}
\end{eqnarray}
From the second integral of (\ref{Rhdef}) and the definition (\ref{Omega}), one gets  
\begin{eqnarray}
{R}'_h&=&\mathcal{H}R_h-\frac{1}{a^3}\,\,\,\,,\nonumber\\
&=& \frac{1}{a^3}\left(\frac{b}{\sqrt{\Omega_V}}-1\right)\,\,\,.\label{manuel16}
\end{eqnarray}
By assuming the dark energy as the dominant component of the universe its conservation law reads 
\begin{eqnarray}\label{manuel17}
{\rho}'_{V}+3\mathcal{H}\left(\rho_V+p_V\right)=0\,\,\,.
\end{eqnarray}
From (\ref{manuel12}) and (\ref{manuel15}), we easily write the derivative of the holographic energy density with respect to $\tau$ as
\begin{eqnarray}\label{manuel18}
{\rho}'_V=-\frac{2}{a^3R_h}\left(\frac{b}{\sqrt{\Omega_V}}-1\right)\rho_V\,\,\,,
\end{eqnarray}
from which, using (\ref{manuel17}), one gets
\begin{eqnarray}\label{manuel19}
\omega_V=-\left(\frac{1}{3}+\frac{2\sqrt{\Omega_V}}{3b}\right)\,\,\,,
\end{eqnarray}
which is the same expression as that obtained by using directly the cosmic time.  In the future, the holographic dark energy will fill the universe such that  $\Omega_V\rightarrow 1$. Then, 
for $b>1$, one gets  $\omega_V>-1$ and the universe will end up in a quintessence-like phase; for $b=1$ the universe will fail into a de Sitter phase, and for $b<1$, the universe falls into a phantom phase and the equation of state crosses $-1$. Therefore, it is conclusive that the parameter $b$ plays an important role when pinpointing the evolutionary nature of the holographic dark energy.\par
Now, one can rewrite the equations (\ref{manuel10}-\ref{manuel11})  in order to point out the correspondence of the  holographic dark energy from the algebraic function  $g(T)$, 
\begin{eqnarray}
  3a^6\mathcal{H}^2&=&\rho+\rho_V\,\,\,,\quad \quad \rho_V=-6a^6\mathcal{H}^2g_T-\frac{1}{2}g\,\,\,,\label{jonas1}\\
-a^6\left(9\mathcal{H}^2+2\mathcal{H}'\right)&=& p+p_V\,\,\,,\quad \quad p_V=  -24a^{12}\mathcal{H}^2\left(3\mathcal{H}^2+\mathcal{H}'\right)g_{TT}+
2a^6\left(6\mathcal{H}^2+\mathcal{H}'\right)g_T+\frac{1}{2}g\,\,\,.\label{jonas2}
\end{eqnarray}
By Combining (\ref{jonas1}) and (\ref{jonas2}), one gets   
\begin{eqnarray}\label{manuel20}
\rho_V+p_V = -24a^{12}\mathcal{H}^2\left(3\mathcal{H}^2+\mathcal{H}'\right)g_{TT}+
2a^6\left(3\mathcal{H}^2+\mathcal{H}'\right)g_T\,\,\,.
\end{eqnarray} 
With the use of (\ref{manuel19}) the left hand side of (\ref{manuel20}) can be rewritten as
\begin{eqnarray}\label{manuel21}
\left(1+\omega_V\right)\rho_V=-6Qa^6\mathcal{H}g_T-\frac{Q}{2}g\,\,\,,\quad\quad Q=\frac{2}{3}\left(1-\frac{\sqrt{\Omega_V}}{b}\right)
\end{eqnarray} 
Thus, from  the right hand sides of (\ref{manuel21}) and (\ref{manuel20}), one gets the following equation 
\begin{eqnarray}
-24a^{12}\mathcal{H}^2\left(3\mathcal{H}^2+\mathcal{H}'\right)g_{TT}+
2a^6\left[3(1+Q)\mathcal{H}^2+\mathcal{H}'\right]g_T+\frac{Q}{2}g=0\;.\label{jonas12}
\end{eqnarray}
Now we have to determine the algebraic function $g(T)$ according to the holographic dark energy. To this end, we assume the power-law scale factor in terms of the cosmic time 
\begin{eqnarray}\label{tscale}
a(t)=\left(\frac{t}{t_0}\right)^\alpha\;\;, \quad\quad \alpha=\frac{2}{3(1+\omega_{eff})}
\end{eqnarray} 
where $\alpha$ is a parameter according to what the stage of the universe can be specify, depending on the effective parameter of EoS $\omega_{eff}=p_{eff}/\rho_{eff}$ ($p_{eff}=p+p_V$ and $\rho_{eff}=\rho+\rho_V$) and $t_0$ the today value of the cosmic time. In terms of $\tau$, the scale factor takes the following expression\footnote{Here we have used the (\ref{ttau})}
\begin{eqnarray}\label{scaletau}
a[t(\tau)]=\left(\frac{\tau}{\tau_0}\right)^\sigma\;,\quad \quad \sigma=\frac{\alpha}{1+3\alpha}\;,\quad \tau_0=\frac{t_0}{1+3\alpha}\;.
\end{eqnarray}
Thus the parameter the torsion scalar (\ref{manuel8}) takes the following form
\begin{eqnarray}
T=-6\frac{\sigma^2}{\tau_0^2}\left(\frac{\tau}{\tau_0}\right)^{2(3\sigma-1)}\;,
\end{eqnarray}
such that the scale factor, the unimodular Hubble parameter and its first derivative are expressed in terms of torsion scalar as
\begin{eqnarray}
a=a_*\left(-T\right)^{\frac{\sigma}{2(3\sigma-1)}}\;\;, \mathcal{H}=\mathcal{H}_*\left(-T\right)^{\frac{1}{2(1-3\sigma)}}\;,\quad\quad \mathcal{H}'=\mathcal{H}_\bullet\left(-T\right)^{\frac{1}{1-3\sigma}}\label{steph1}\\
a_*=\left(\frac{\tau_0}{\sigma\sqrt{6}}\right)^{\frac{\sigma}{3\sigma-1}}\;,\quad 
\mathcal{H}_*=\frac{\sigma}{\tau_0}\left(\frac{\tau_0}{\sigma\sqrt{6}}\right)^{\frac{1}{1-3\sigma}}\;,\quad\quad \mathcal{H}_\bullet=-\frac{\sigma}{\tau^2_0}\left(\frac{\tau_0}{\sigma\sqrt{6}}\right)^{\frac{2}{1-3\sigma}}
\end{eqnarray}
By using the expressions in (\ref{steph1}), the equation (\ref{jonas12}) gives rive to the following differential equation
\begin{eqnarray}
T^2g_{TT}+\lambda_1Tg_T+\lambda_2g=0\;,\label{eqdiffQ}\\
\lambda_1=\frac{\mathcal{H}_\bullet+3(1+Q)\mathcal{H}_*^2}{12a_*^6\mathcal{H}_*^2\left(\mathcal{H}_\bullet+3\mathcal{H}_*^2   \right)}= \frac{3\sigma(Q+1)-1}{2(3\sigma-1)}\;\;,\quad 
\lambda_2=-\frac{Q}{48a_*^{12}\mathcal{H}_*^2\left(\mathcal{H}_\bullet+3\mathcal{H}_*^2   \right)}=-\frac{3Q\sigma}{4(3\sigma-1)}.
\end{eqnarray}
It is important to point out that the resolution of the differential equation (\ref{eqdiffQ}) at this stage is not possible because of the explicit expression of the parameter $Q$ is not known. Remember that $Q$ depends essentially on $\Omega_V$ and this later depends on the future event horizon $R_h$. Let us look for $R_h$ determining its explicit time dependent expression. Then, making use of (\ref{scaletau}) and considering that $R_h$ must vanish for large value of $\tau$, one imposes the condition $\sigma>1/4$ and gets
\begin{eqnarray}\label{Omegasigma}
R_h=\frac{\tau_0}{4\sigma-1}\left( \frac{\tau}{\tau_0}\right)^{1-3\sigma}\;,\quad \Omega_V=\frac{b^2(4\sigma-1)^2}{\sigma^2}\;,\quad Q=\frac{2(1-3\sigma)}{3\sigma}\;,
\end{eqnarray}
such that
\begin{eqnarray}
\lambda_1=-\frac{1}{2}\;\quad\quad\mbox{and}\quad\quad\lambda_2=\frac{1}{2}\;.\label{lambdasigma}
\end{eqnarray}

Now is clear that the parameter $Q$ is constant, and the general solution of (\ref{eqdiffQ}) reads
\begin{eqnarray}
g(T)=C_1\left(-T\right)+C_2\left(-T\right)^{\frac{1}{2}}\;
\end{eqnarray}
such that the  $f(T)$ holographic dark energy model in the unimodular context reads
\begin{eqnarray}
f(T)=\left(1-C_1\right)T+C_2\left(-T\right)^{\frac{1}{2}}\;,
\end{eqnarray}  
where $C_1$ and $C_2$ are integration constants. For more consistency, the constants have been to be determined and to to do so, we impose the initial conditions, assuming the assumption according to what, at present time, the holographic model must recover the usual $\Lambda$CDM  one, that is
\begin{eqnarray}
\left(f\right)_{t=t_0}=T_0-2\Lambda\;\;,\quad\quad \left(\frac{df}{dt}\right)_{t=t_0}=\left(\frac{dT}{dt}\right)_{t=t_0}\;,\label{icond}
\end{eqnarray}
where the subscript $t_0$ and $T_0$ denote the present time and the related value of the torsion scalar, respectively.\par
Making use of the initial conditions (\ref{icond}), one gets
\begin{eqnarray}
C_1=-\frac{2\Lambda}{T_0}\;,\quad\quad C_2=-4\Lambda (-T_0)^{-\frac{1}{2}}\:,
\end{eqnarray}
such that the algebraic unimodular holographic dark energy model reads
\begin{eqnarray}
f(T)=\left(1+\frac{2\Lambda}{T_0}\right)T-4\Lambda\left(\frac{T}{T_0}\right)^{\frac{1}{2}}\;.\label{genemodel}
\end{eqnarray}

\section{Fitting the reconstructed unimodular holographic dark energy model with observational data}\label{sec4}

Here let us cast Eq. (\ref{jonas1}) in the following form
\begin{eqnarray}
\frac{B(z)}{B_0}=\left[\Omega_{m0}(1+z)^{3(1+\omega)}+(1-\Omega_{m0})(1+z)^{3(1+\omega_V)}\right]^{\frac{1}{2}}\;,
\end{eqnarray}
where $B(z)=a^3(\tau(z))\mathcal{H}(\tau(z))$ is the equivalent of the cosmic time dependent Hubble parameter $H(t(z))$, and $E_0$ its today value. On the same way we shall compute the decelerated parameter $q=-a\ddot{a}/\dot{a}^2$ as
\begin{eqnarray}
q&=&-4-\frac{\mathcal{H}'}{\mathcal{H}^2}\\
&=& -4+\frac{3}{2}\frac{\Omega_{m0}(3+\omega)(1+z)^{3(3+\omega)}+(1-\Omega_{m0})(3+\omega_V)(1+z)^{3(3+\omega_V)}}{\Omega_{m0}(1+z)^{3(3+\omega)}+(1+\Omega_{m0})(1+z)^{3(3+\omega_V)}}\;
\end{eqnarray}
and the effective parameter of EoS $\omega_{eff}$ as
\begin{eqnarray}
\omega_{eff}&=&-3-\frac{2\mathcal{H}'}{3\mathcal{H}^2}\\
&=& -3+\frac{\Omega_{m0}(3+\omega)(1+z)^{3(3+\omega)}+(1-\Omega_{m0})(3+\omega_V)(1+z)^{3(3+\omega_V)}}{\Omega_{m0}(1+z)^{3(3+\omega)}+(1+\Omega_{m0})(1+z)^{3(3+\omega_V)}}\;.
\end{eqnarray}
In the above expressions we have used the relation $\mathcal{H}'=-(1+z)\mathcal{H}d\mathcal{H}/dz$.\par
Our challenge hear is to compare the background expansion in unimodular holographic dark energy model with observational data. The task if to investigate whether the cosmology provide by the reconstructed model accords the available background data. To do so, we use the typical $H(z)$ data coming from the cosmic chronometers. Measuring $H(z)$ and for instance $B(z)$, using the differential age of the universe therefore circumvents the limitations associated with the use of the integrated histories. The best cosmic chronometer concerns the galaxies evolving passively on a time scale much longer than their age difference. In this work we assume the ordinary matter content as a pressure-less fluid. The $H(z)$ data to be used here is in the range $0<z<2$. The current value of the Hubble parameter we assume here is $H_0=70 km/s/Mpc$ and then the value of $\Omega_{m0}$ can be consequently determined through the present conditions applied to the Eq. (\ref{manuel9}), using (\ref{genemodel}) and leading to $\Omega_{m0}=0.3$, the cosmological constant being $\Lambda=0.7$. Our result should be compare with the standard well known $\Lambda CDM$ and the pure matter dominated Einstein-de Sitter models, for which one has, respectively
\begin{eqnarray}
E_{\Lambda CDM}(z)&=&E_0\left[\Omega_{m0}(1+z)^3+\Lambda\right]^{\frac{1}{2}}\\
E_{EdS}(z)&=&E_0(1+z)^{\frac{3}{2}}
\end{eqnarray}
By making use of $\Omega_V$ at (\ref{Omegasigma}), the expression (\ref{manuel19}) becomes
\begin{eqnarray}\label{dernier}
\omega_V=-\frac{1}{3}\left[1+\frac{2(4\sigma-1)}{\sigma}\right]\;.
\end{eqnarray}
Due to the fact that $\sigma>1/4$, one get $\omega_V< -1/3$. Moreover, from (\ref{dernier}), one gets $\sigma=2/(3\omega_V+9)$, which, within the condition $\sigma>1/4$, yields $-3<\omega_V< -1/3$ as a crucial condition. Then for any curve to be plotted, one has to refer to the previous condition about $\omega_V$. However it has to be pointed out that for $\omega=-1$, corresponding to $\sigma=1/3$, the $\Lambda CDM$ model is recovered. Then, except the value $-1$, we fix three other values for the parameter $\omega_V$ in the range $-3<\omega_V< -1/3$ and plot the evolution the Hubble parameter, the deceleration parameter and the effective parameter of EoS versus the red-shift, that is $H(z)$, $q(z)$ and $\omega_{eff}(z)$. 
\begin{figure}[h]
\centering
\begin{tabular}{rl}
\includegraphics[width=7cm, height=7cm]{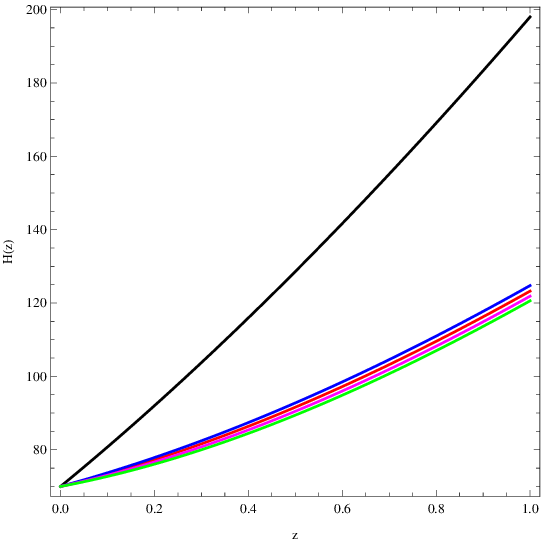}&
\includegraphics[width=7cm, height=7cm]{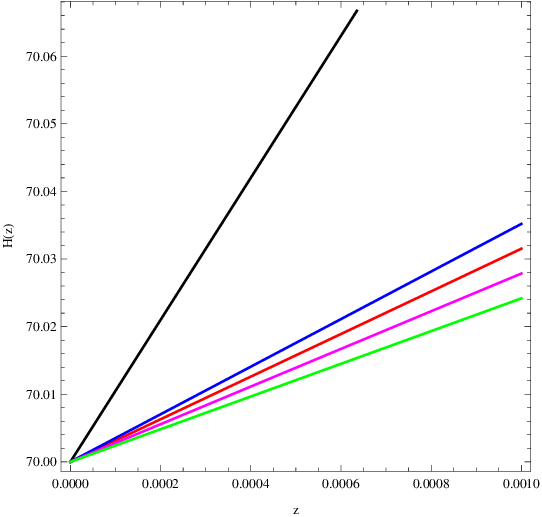}
\end{tabular}
\caption{Left panel: evolution of $H(z)$. Right panel: The inset of the evolution of $H(z)$ amplifying the low-$z$ region. Here we consider $\Omega_{m0}=0.3$, $\Lambda=0.7$ and $H_0=70 km/s/Mpc$. The Black and Red characterize the EdS and $\Lambda CDM$ models, respectively, while the Green, Magenta and Blue are related to the UHDE model for $\omega_V=-1.1, -1.05, -0.95$      }
\label{fig1}
\end{figure}

\begin{figure}[h]
\centering
\begin{tabular}{rl}
\includegraphics[width=7cm, height=7cm]{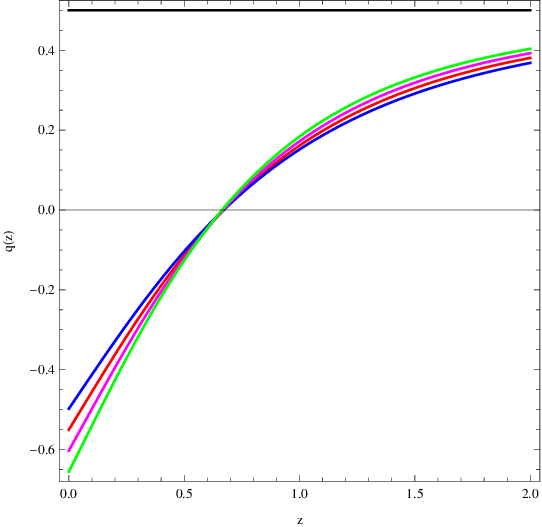}&
\includegraphics[width=7cm, height=7cm]{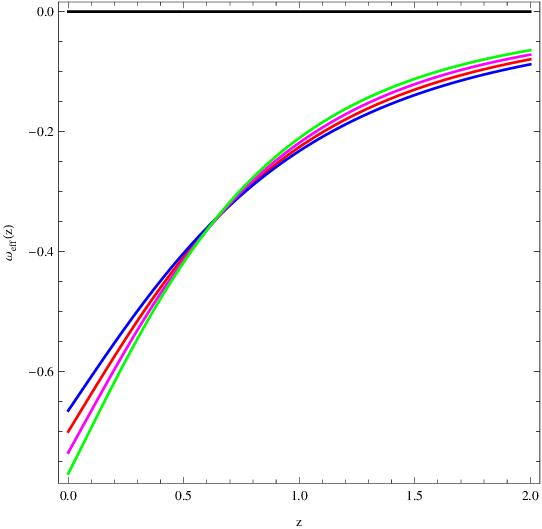}
\end{tabular}
\caption{Left panel: evolution of $q(z)$. Right panel: evolution of $\omega_{eff}(z)$. We use $\Omega_{m0}=0.3$, $\Lambda=0.7$ and $H_0=70 km/s/Mpc$. The Black and Red characterize the EdS and $\Lambda CDM$ models, respectively, while the Green, Magenta and Blue are related to the UHDE model for $\omega_V=-1.1, -1.05, -0.95$}
\label{fig2}
\end{figure}
\begin{figure}[h]
\centering
\begin{tabular}{rl}
\includegraphics[width=7cm, height=7cm]{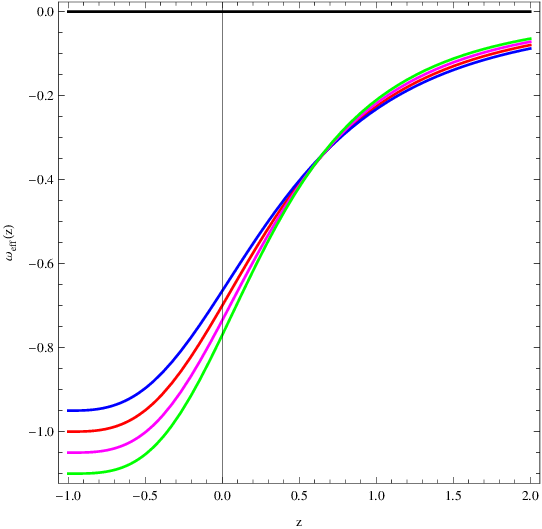}
&\includegraphics[width=7cm, height=7cm]{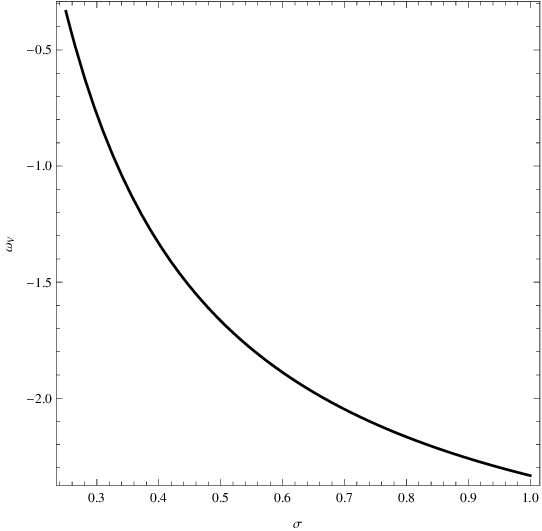}
\end{tabular}
\caption{Left panel: evolution of $\omega_{eff}(z)$. We use $\Omega_{m0}=0.3$, $\Lambda=0.7$ and $H_0=70 km/s/Mpc$. The Black and Red characterize the EdS and $\Lambda CDM$ models, respectively, while the Green, Magenta and Blue are related to the UHDE model for $\omega_V=-1.1, -1.05, -0.95$. Left panel: evolution of $\omega_V$ versus $\sigma$.}
\label{fig3}
\end{figure}

From Fig.\;\ref{fig1} it appears that the EdS, $\Lambda CDM$ and the UHDE models, the behavior the Hubble parameter obeys the standard accelerated expansion feature, that is, for an accelerated expansion of the universe it hopped to have a decreasing rate. However, two other important aspects have to be analyzed in order to point out the viability of the reconstructed UHDE model, that is, the transition from the decelerated to the accelerated phases of the universe through the decelerated parameter $q(z)$, and the crossing of $-1$ by the effective EoS $\omega_{eff}$ predicting the possibility of having finite time singularities in the future.  As well known the EdS model should note provide the transition phase and the universe should live only the decelerated expansion ($q(z)>0$) as shown by the black curve in the left panel of Fig.\;\ref{fig2}. For the $\Lambda CDM$ model the transition is realized as well shown by the red curved for the $\omega_V=-1.1, 1.05, -0.95$. About the UHDE model the transition is also guaranteed crediting the model as candidate to the viability test. In order to give more consistency to the analysis, the evolution of $\omega_{eff}$ is plotted for the same models as in the previous case. Note in this case that, just  from the higher values of $z$ to the present time, only the EdS model does note present an  unexpected behavior, being always positive, as shown in the right panel of Fig.\;\ref{fig2}; the $\Lambda CDM$ and UHDE models agree with the WMAP9 result ($-1.71<\omega_{eff}<-0.34 $) at $z=0$ but any one of current values of $\omega_V$ does not cross $-1$. Therefore, it is clear that the most important role of $\omega_{eff}$ shall appear when one goes toward the future, that is, allowing the redshift $z$ to possibly reach $-1$. At this stage (see Fig\;\ref{fig3}),  the parameter $\omega_{eff}$ vanishes for the EdS model, being in disagreement with the observational data. Having a look on $\Lambda CDM$ and UHDE models, one sees that $\omega_{eff}$ will never cross $-1$ for $\Lambda CDM$ (for which the universe end up with $\omega_{eff}=-1$) and for UHDE model within $\omega_V=-0.95$, while the transition from the quintessence to the phantom phase is well realized for UHDE model within $\omega_V=-1.1, 1.05$. In accordance with the analysis of the $q(z)$ and $\omega_{eff}$, the UHDE model passes the $H(z)$ data test, at least for the range $-1.10\leq\omega_V\leq -1.05$. For more precision we add the evolution the $\omega_V$ in terms of the parameter $\sigma$ at the right panel of Fig.\;\ref{fig3}, where the graph shows a decreasing behavior of $\omega_V$. \par
In order to complete the viability of the reconstructed UHDE model, studying its stability against linear geometrical and matter perturbations. The present this analysis in the coming section where the de Sitter and power-law solutions are considered.

\newpage
\section{Stability of cosmological solutions}\label{sec5}
In this section we explore the stability feature of the reconstructed model. This study requires to to introduce homogeneous and isotropic perturbations around the model. Because of the form of the geometrical part of the field equation (\ref{manuel9}-\ref{manuel10}), it is useful to assume the scale factor $a_b(\tau)$ satisfying these equations.  We now consider small deviations from the scale factor and the ordinary energy density in terms of perturbation functions as
\begin{eqnarray}
a(\tau)=a_b(\tau)\left[1+\delta(\tau)\right]\;,\quad\quad \rho(\tau)=\rho_b(\tau)\left[1+\delta_m(\tau)\right].
\end{eqnarray} 
Here $\delta(\tau)$ and $\delta_m(\tau)$ denote the geometrical and matter perturbation functions, respectively. For our purpose in this work,  we assume linear perturbation and then one gets
\begin{eqnarray}
\mathcal{H}(\tau)=\mathcal{H}_b(\tau)+\delta'(\tau)\;.
\end{eqnarray}
Moreover, we propose to expand the algebraic function $f(T)$ about the value of the torsion scalar in the background, namely $T_b=-6a^6\mathcal{H}^2$, as
\begin{eqnarray}
f(T)=f(T_b)+f_T(T_b)\left(T-T_b\right)+\frac{1}{2}f_{TT}(T_b)\left(T-T_b\right)^2+\mathcal{O}^3, \;,\label{dlim}
\end{eqnarray}
where $\mathcal{O}^3$ represents the terms of higher power of $T$ being neglected. Making use of (\ref{dlim}), the equation (\ref{manuel9}) takes the following form
\begin{eqnarray}
6a_b^6\mathcal{H}_b\left(f_T-12a_b^6\mathcal{H}^2_bf_{TT}\right)\delta'+18a_b^6\mathcal{H}^2_b\left(f_T-12a_b^6\mathcal{H}^2_bf_{TT}\right)\delta=\rho_b\delta_m\;,\label{pertfried1}
\end{eqnarray}
where $\rho_b(t)$ can be obtained by solving the matter equation of continuity 
\begin{eqnarray}
\rho'+3\mathcal{H}(1+\omega)\rho=0\;,\label{mateqcont}
\end{eqnarray}
 getting 
\begin{eqnarray}
\rho_b(\tau)=\rho_0\exp{\left\{-3(1+\omega)\int \mathcal{H}_b(\tau)d\tau\right\}}\;.\label{soleqcont}
\end{eqnarray}
The equation (\ref{pertfried1}) shows a relationship between the geometrical and matter perturbation functions, but it is quite clear that it is not sufficient for determining each perturbation function. Therefore we also perturb the matter equation of continuity (\ref{mateqcont}), obtaining
\begin{eqnarray}
\delta'_m+3\left(1+\omega\right)\delta'=0\;.\label{perteqcont}
\end{eqnarray}
By integrating (\ref{perteqcont}) and withdrawing the additive constant, for simplicity, one gets
\begin{eqnarray}
\delta_m+3\left(1+\omega\right)\delta=0\;.\label{solperteqcont}
\end{eqnarray}
By extracting $\delta_m$ from (\ref{solperteqcont}) and injecting in (\ref{pertfried1}), one gets
\begin{eqnarray}
6a_b^6\mathcal{H}_b\left(f_T-12a_b^6\mathcal{H}^2_bf_{TT}\right)\delta'+\left\{18a_b^6\mathcal{H}^2_b\left(f_T-12a_b^6\mathcal{H}^2_bf_{TT}\right)+3\rho_0(1+\omega)\exp{\left[-3(1+\omega)\int \mathcal{H}_b(\tau)d\tau\right]}\right\}\delta=0\;,
\end{eqnarray}
whose general solution reads
\begin{eqnarray}
\delta(\tau)=K\exp\left\{-\int \frac{S(\tau)}{R(\tau)}d\tau\right\}\;,
\end{eqnarray}
and consequently 
\begin{eqnarray}
\delta_m(\tau)=-3K(1+\omega)\exp\left\{-\int \frac{S(\tau)}{R(\tau)}d\tau\right\}\;,
\end{eqnarray}
with $K$ an integration constant and
\begin{eqnarray}
R(\tau)&=&6a_b^6(\tau)\mathcal{H}_b(\tau)\left[f_T(\tau)-12a_b^6(\tau)\mathcal{H}^2_b(\tau)f_{TT}(\tau)\right]\;,\\
S(\tau)&=&18a_b^6(\tau)\mathcal{H}^2_b(\tau)\left[f_T(\tau)-12a_b^6(\tau)\mathcal{H}^2_b(\tau)f_{TT}(\tau)\right]+3\rho_0(1+\omega)\exp{\left[-3(1+\omega)\int \mathcal{H}_b(\tau)d\tau\right]}
\end{eqnarray}
In order to find the explicit $\tau$ dependent expression of the $\delta$, one has to fix the $\tau$ dependent expression of the scale factor; to do so, we will assule the de Sitter and power-law cosmological solutions.
\subsection{Stability of de Sitter solutions}
Here we first assume the cosmic time dependent expression of the de Sitter solution as
\begin{eqnarray}
a(t)=a_0e^{H_0t}\;,
\end{eqnarray}
where $H_0$ is a positive constant, more precisely the present value of the Hubble parameter. In this way, the $\tau$ dependent de Sitter solution, using (\ref{ttau}), reads
\begin{eqnarray}
a_b(\tau)=\left(3H_0\right)^{1/3}\tau^{1/3}\Longrightarrow \mathcal{H}_b(\tau)=\frac{1}{3\tau}\Longrightarrow
T_b=-6a_b^6\mathcal{H}_b^2=-6H_0^2\;,
\end{eqnarray}  
such that 
\begin{eqnarray}
R(\tau)&=&6\left(3H^2_0-\Lambda\right)\tau\;,\\
S(\tau)&=&6\left(3H^2_0-\Lambda\right)+3\frac{\rho_0(1+\omega)}{\tau^{1+\omega}}\;.
\end{eqnarray}
Then, one gets the following solutions for the perturbation functions
\begin{eqnarray}
\delta(\tau)=\frac{K}{\tau}\exp\left\{\frac{\rho_0}{2(3H^2_0-\Lambda)\tau^{1+\omega}}\right\}\;,\\
\delta_m(\tau)=-3\frac{K(1+\omega)}{\tau}\exp\left\{\frac{\rho_0}{2(3H^2_0-\Lambda)\tau^{1+\omega}}\right\}\;.
\end{eqnarray}

\subsection{Stability of power-law solutions}
In this subsection we focus to cosmological time dependent solution, being of the type
\begin{eqnarray}
a(t)=\left(\frac{t}{t_0}\right)^\alpha\Longrightarrow a_b(\tau)=\left(\frac{\tau}{\tau_0}\right)^\sigma\;,
\end{eqnarray}
where $\alpha$, $\sigma$ and $\tau_0$ are the parameters defined at (\ref{tscale}) and (\ref{scaletau}). 
More precisely, we fix $\omega_V=-21/20$ (this value, $-1.05$ refers to the Magenta line in the figures) for which there is a best fit with observational data. With this value, one gets $\sigma=40/117>1/4$.
Thus, one gets
\begin{eqnarray}
\mathcal{H}_b(\tau)\propto\frac{\sigma}{\tau}\Longrightarrow T_b(\tau)\propto -6\frac{\sigma^2}{\tau_0^2}\left(\frac{\tau}{\tau_0}\right)^{2(3\sigma-1)}\;.
\end{eqnarray}
In this case, one gets
\begin{eqnarray}
R(\tau)&=&  \frac{2}{\sigma\tau_0}\left(3\sigma^2-\Lambda\tau_0^2\right)\left(\frac{\tau}{\tau_0}\right)^{6\sigma-1}\;,\\
S(\tau)&=& \frac{6}{\tau_0^2}\left(3\sigma^2-\Lambda\tau_0^2\right)\left(\frac{\tau}{\tau_0}\right)^{2(3\sigma-1)}+3\rho_0\left(1+\omega\right)\tau^{-3\sigma(1+\omega)}
\end{eqnarray}
As in the previous section, we assume ordinary matter and dust, such that $\omega=0$. Therefore, the perturbation functions read
\begin{eqnarray}
\delta(\tau)=\frac{K}{\tau^{3\sigma}}\exp{\left[-\frac{3\rho_0\sigma\tau_0^{6\sigma}}{2(2-9\sigma)(3\sigma^2-\Lambda\tau_0^2)}\frac{1}{\tau^{9\sigma-2}}\right]}\;,\\
\delta_m(\tau)=\frac{-3K}{\tau^{3\sigma}}\exp{\left[-\frac{3\rho_0\sigma\tau_0^{6\sigma}}{2(2-9\sigma)(3\sigma^2-\Lambda\tau_0^2)}\frac{1}{\tau^{9\sigma-2}}\right]}
\end{eqnarray}

\begin{figure}[h]
\centering
\begin{tabular}{rl}
\includegraphics[width=7cm, height=7cm]{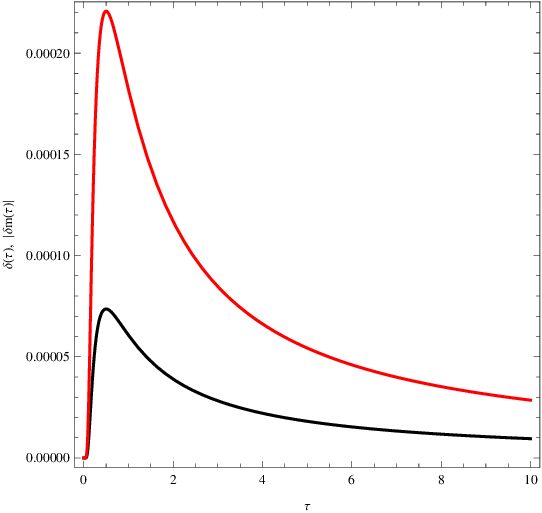}&
\includegraphics[width=7cm, height=7cm]{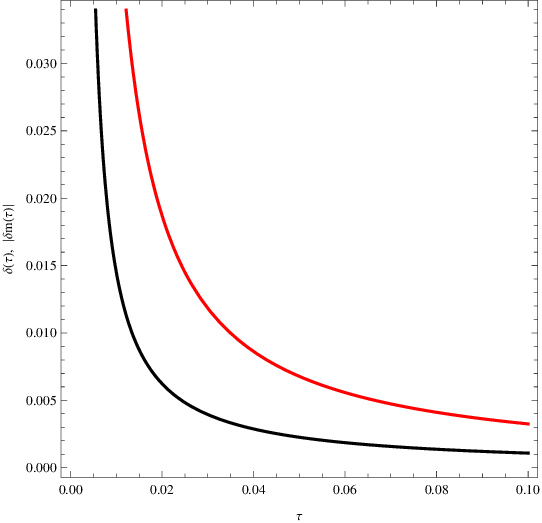}
\end{tabular}
\caption{Left panel: evolution of $\delta(\tau)$ (black line) and $\delta_m(\tau)$ (red line) for the de Sitter solutions. Right panel: evolution of $\delta(\tau)$ (black line) and $\delta_m(\tau)$ (red line) for the de Sitter solutions. We use $\Omega_{m0}=0.3$, $\Lambda=0.7$, $H_0=70 km/s/Mpc$, $\sigma=40/117$ and $t_0=1/H_0$.}
\label{fig4}
\end{figure}
The Fig.\;\ref{fig4} presents the evolutions of the perturbation functions for both de sitter  and power-law solutions.  Due to the time parametrization, passing from the cosmic time to $\tau$, the gravitational perturbation is operated directly on the scale factor for simplicity yielding a linear dependence of the $\delta(\tau)$ on the matter perturbation function $\delta_m(\tau)$, such that when the for positive value of one of the them, the second is negative. With a positive integration constant $K$, one gets $\delta(\tau)>0$ and $\delta_m(\tau)<0$. Then, we plot the evolution of $\delta(\tau)$ and $|\delta_m(\tau)|$. The question to be asked is to known if the cosmic time $t$ and the parametrized $\tau$ obey the same rate of evolution. Note that $d\tau/dt=a^3>0$ , this means that as the cosmic time evolves, $\tau$ also evolves, allowing to conclude from Fig.\;\ref{fig4} that as the cosmic time evolves and for its large values the perturbation functions go toward vanishing values. Thus the reconstructed unimodular holographic dark energy model can be considered as stable.

\section{Conclusion}\label{sec6}
We explore unimodular $f(T)$ theory of gravity by reconstructing the related model able to drive the cosmological physical properties of the holographic dark energy. The unimodular notion appears by fixing the determinant of the tetrad to a number, and  in this paper it is set to $1$. We look for the viability of the reconstructed model by fitting it with the cosmological $H(z)$ data and also study its stability for consistent analysis.\par
About the $H(z)$ test we focus to three fundamental parameters, namely, the Hubble parameter, the decelerated parameter and the effective equation of state parameter. Attention has been attached to the $\Lambda CDM$ model, the Einstein de Sitter model and the reconstructed holographic dark energy model, this later, essentially depending on the parameter of equation of state related to the holographic dark energy, constrained to $-3<\omega_V<-1/3$. Then we consider three values of the parameter, namely, $\omega_V=-1.10, -1.05, -0.95$, and plot the evolution of the Hubble parameter, the decelerated parameter and the parameter of effective equation of state, for the Einstein de sitter model, $\Lambda CDM$ model and the reconstructed holographic dark energy model for $\omega_V=-1.10, -1.05, -0.95$. The evolutions of the Hubble parameter versus red shift $z$ for the considered models reflect the expected behavior in concordance with the expansion of the universe. About the decelerated parameter $q(z)$, except the Einstein de Sitter for which the universe in always decelerating, the transition from the decelerated to the accelerated phase is realized for the $\Lambda CDM$ and the holographic dark energy models. On the other hand, looking for the effective, also except the EdS model, the $\Lambda CDM$ and HDE models are in agreement with the WMAP9 result at the present time ($z=0$). However, as the red-shift evolves toward $z=-1$, i.e in future, only the HDE model provides the transition from the quintessence to the phantom, that is, crossing $\omega_{eff}=-1$, for $\omega_{V}=-1.10,-1.05$. We then conclude that at least in the range $-1.10\leq \omega_V\leq -1.05$ the unimodular holographic dark energy passes the $H(z)$ test.\par
Moreover, we submit the reconstructed model to gravitational and matter linear perturbations, considering the de Sitter and power-law solutions to the scale factor, and fixing $\omega_V$  to $-1.05$. The results show in both cases that as the time evolves, the perturbation functions decrease and go toward almost vanishing values. Thus we conclude that the reconstructed model is stable against the considered linear perturbations, and should be assumed as alternative viable model to the TT.

\vspace{0.25cm}
{\bf Acknowledgement:}


\begin{thebibliography}{17}
\addcontentsline{toc}{chapter}{Bibliographie}
\bibitem{1dewang} J. D. Bekenstein, Phys. Rev. D {\bf 7}, 2333-2346 (1973).
\bibitem{2dewang} S. W. Hawking, Commun. Math. Phys. {\bf 43}, 199-220 (1975).
\bibitem{3dewang} G. 't Hooft, arXiv:gr-qc/9310026.
\bibitem{5dewang} J. M. Maldacena, Int. J. Theor. Phys. {\bf 38}, 1113-1133 (1999) .
\bibitem{6dewang} H. Liu, K. Rajagopal, U. A. Wiedemann, JHEP {\bf 03}, 066 (2007).
\bibitem{7dewang} S. A. Hartnoll, Class. Quant. Grav. {\bf 26}, 224002 (2009).
\bibitem{8dewang} T. Takayanagi,  Class. Quant. Grav. {\bf 29}, 153001 (2012).
\bibitem{9dewang} A. Strominger, JHEP {\bf 10}, 034 (2001).
\bibitem{1depurba} S. M. Carroll, Living Rev. Rel. {\bf4}, 1 (2001).
\bibitem{2depurba} T. Padmanabhan, Phys. Rept. {\bf 380}, 235 (2003).
\bibitem{3depurba} B. Ratra and P. J. E. Peeble, Phys. Rev. D 37, 3406 (1988);
\bibitem{int} I. Zatev, L. Wang and P. J. Steinhardt, Phys. Rev. Lett. {\bf 82}, 896 (1999);
\bibitem{int} M. Sahlen, A. R. Liddle and D. Parkinson, Phys. Rev. D {\bf 75}, 023502 (2007);
\bibitem{int} N. Banerjee and S. Das, Gen. Rel. Grav., {\bf 37}, 1695 (2005);
\bibitem{int} R. J. Scherrer and A. A. Sen, Phys. Rev. D {\bf 77}, 083515 (2008);
R. J. Scherrer and A. A. Sen, Phys. Rev. D {\bf 78}, 067303 (2008);
\bibitem{int} T. Chiba, Phys. Rev. D {\bf 79}, 083517 (2009); T. Chiba, Phys. Rev. D {\bf 80}, 109902 (2009);
\bibitem{3depurbaf} G. Gupta, R. Rangarajan and A. A. Sen, Phys. Rev. D {\bf 92}, 123003 (2015).
\bibitem{4depurba} R. R. Caldwell, Phys. Lett. B {\bf 545}, 23 (2002).
\bibitem{5depurba} G. W. Gibbons, Phys. Lett. B {\bf 537}, 1 (2002);
\bibitem{int} T. Padmanabhan, Phys. Rev. D {\bf 66}, 021301 (2002);
\bibitem{int} J. S. Bagla, H. K. Jassal and T. Padmanabhan, Phys. Rev. D {\bf 67}, 063504 (2003);
\bibitem{5depurbaf} E. J. Copeland, M. R. Garousi, M. Sami and S. Tsujikawa, Phys. Rev. D {\bf 71}, 043003 (2005).
\bibitem{6depurba} A. Y. Kamenshchik, U. Moschella and V. Pasquier, Phys. Lett. B {\bf 511}, 265 (2001);
\bibitem{6depurbaf} M. C. Bento, O. Bertolami, A. A. Sen, Phys. Rev. D {\bf 66}, 043507 (2002); M. C. Bento, O. Bertolami and A. A. Sen, Phys. Rev. D {\bf 67}, 063003 (2003).
\bibitem{fRi} S. Nojiri, S.D. Odintsov, V.K. Oikonomou, arXiv:1710.07838 [gr-qc]; 
S.D. Odintsov, V.K. Oikonomou,  arXiv:1710.01226 [gr-qc]; S.D. Odintsov, V.K. Oikonomou, L. Sebastiani,  Nucl.Phys. B {\bf 923} (2017) 608-632; Andrea Addazi, Shin'ichi Nojiri, Sergei Odintsov, Phys.Rev. D {\bf 95} (2017) no.12, 124020.
\bibitem{fR} S.D. Odintsov, V.K. Oikonomou, Phys.Rev. D {\bf 94} (2016) no.12, 124026; 
S.D. Odintsov, V.K. Oikonomou, Phys.Rev. D {\bf 94} (2016) no.4, 044012; D.J. Brooker, S.D. Odintsov, R.P. Woodard, Nucl.Phys. B {\bf 911} (2016) 318-337; Kazuharu Bamba, Sergei D. Odintsov, Emmanuel N. Saridakis,  Mod.Phys.Lett. A {\bf 32} (2017) no.21, 1750114.
\bibitem{fR} S. Nojiri, S.D. Odintsov, V.K. Oikonomou, DOI: 10.1142/S0217732316501728, arXiv:1605.00993 [gr-qc]; Sebastian Bahamonde, S.D. Odintsov, V.K. Oikonomou, Matthew Wright,  Annals Phys. {\bf 373} (2016) 96-114; S.D. Odintsov, V.K. Oikonomou, Phys.Rev. D{\bf 92} (2015) no.12, 124024. Salvatore Capozziello, Mariafelicia De Laurentis, Ruben Farinelli, Sergei D. Odintsov,  Phys.Rev. D {\bf 93} (2016) no.2, 023501.

\bibitem{fR} Artyom V. Astashenok, Salvatore Capozziello, Sergei D. Odintsov, Phys.Lett. B{ \bf 742} (2015) 160-166; Shin'ichi Nojiri, Sergei D. Odintsov, Astrophys.Space Sci. {\bf 357} (2015) no.1, 39; S.D. Odintsov, V.K. Oikonomou, Phys. Rev. D {\bf 90} (2014) no.12, 124083; Kazuharu Bamba, Shin'ichi Nojiri, Sergei D. Odintsov, Diego S\'aez-G\'omez Phys.Rev. D {\bf 90} (2014) 124061. Artyom V. Astashenok, Salvatore Capozziello, Sergei D. Odintsov, Astrophys.Space Sci. {\bf 355} (2015) no.2, 333-341. 

\bibitem{fR} Shin'ichi Nojiri, Sergei D. Odintsov, Phys.Lett. B {\bf 735} (2014) 376-382; Andrey N. Makarenko, Sergei D. Odintsov, Gonzalo J. Olmo, Phys.Lett. B {\bf 734} (2014) 36-40; R. Farinelli, M. De Laurentis, S. Capozziello, S.D. Odintsov,  Mon. Not. Roy. Astron. Soc. {\bf 440} (2014) 2909-2915; Artyom V. Astashenok, Salvatore Capozziello, Sergei D. Odintsov, JCAP {\bf 1312} (2013) 040.
\bibitem{fR} E. Elizalde, S.D. Odintsov, L. Sebastiani, S. Zerbini, Eur. Phys. J. C {\bf 72} (2012) 1843; Kazuharu Bamba, Shin'ichi Nojiri, Sergei D. Odintsov, Phys. Rev. D {\bf 85} (2012) 044012; Peter K.S. Dunsby, Emilo Elizalde, Sergei Odintsov, Diego Saez Gomez Phys. Rev. D {\bf 82} (2010) 023519; Shin'ichi Nojiri, Sergei D. Odintsov, Diego Saez-Gomez, Phys.Lett. B {\bf 681} (2009) 74-80.
\bibitem{fRf} Shin'ichi Nojiri, Sergei D. Odintsov, Phys. Lett. B {\bf 659} (2008) 821-826; 
Shin'ichi Nojiri, Sergei D. Odintsov, Phys. Rev. D {\bf 74} (2006) 086005; Guido Cognola, Emilio Elizalde, Shin'ichi Nojiri, Sergei D. Odintsov, Sergio Zerbini, JCAP {\bf 0502} (2005) 010; Salvatore Capozziello, S. Nojiri, S.D. Odintsov, A. Troisi, Phys. Lett. B {\bf 639} (2006) 135-143; G. Cognola, E. Elizalde, S.D. Odintsov, P. Tretyakov, S. Zerbini, Phys. Rev. D {\bf 79} (2009) 044001; Shin'ichi Nojiri, Sergei D. Odintsov, TSPU {\bf 110} (2011) 7-19; R.D. Boko, M.J.S. Houndjo, J. Tossa, Int. J. Mod. Phys. D {\bf 25} (2016) no.10, 1650098; M.J.S. Houndjo, A.V. Monwanou, Jean B.Chabi Orou, Int. J. Mod. Phys. D {\bf 20} (2011) 2449-2469.
\bibitem{fGi} J. Haro, A.N. Makarenko, A.N. Myagky, S.D. Odintsov, V.K. Oikonomou, Phys. Rev. D{\bf 92} (2015) no.12, 124026; 
Salvatore Capozziello, Andrey N. Makarenko, Sergei D. Odintsov, Phys. Rev. D{\bf 87} (2013) no.8, 084037;
Kazuharu Bamba, Sergei D. Odintsov, Lorenzo Sebastiani, Sergio Zerbini, Eur. Phys. J. C{\bf 67} (2010) 295-310; Guido Cognola, Emilio Elizalde, Shin'ichi Nojiri, Sergei D. Odintsov, Sergio Zerbini, Eur. Phys. J. C {\bf 64} (2009) 483-494;
\bibitem{fGf} Guido Cognola, Emilio Elizalde, Shin'ichi Nojiri, Sergei D. Odintsov, Sergio Zerbini, Phys. Rev. D {\bf 73} (2006) 084007; Shin'ichi Nojiri, Sergei D. Odintsov, O.G. Gorbunova, J.Phys. A {\bf 39} (2006) 6627-6634; Shin'ichi Nojiri, Sergei D. Odintsov, Phys.Lett. B {\bf 631} (2005) 1-6; Shin'ichi Nojiri, Sergei D. Odintsov, Misao Sasaki, Phys. Rev. D {\bf 71} (2005) 12350;  
M.E. Rodrigues, M.J.S. Houndjo, D. Momeni, R. Myrzakulov, Can. J. Phys. {\bf 92} (2014) 173-176; 
M.J.S. Houndjo, M.E. Rodrigues, D. Momeni, R. Myrzakulov, Can. J. Phys. {\bf 92} (2014) no.12, 1528-1540.

\bibitem{fRTi} Tiberiu Harko, Francisco S.N. Lobo, Shin'ichi Nojiri, Sergei D. Odintsov, Phys.Rev. D {\bf 84} (2011) 024020; M.J.S. Houndjo, Int. J. Mod. Phys. D{\bf 21} (2012) 1250003; M.J.S. Houndjo, Oliver F. Piattella, Int. J. Mod. Phys. D{\bf 21} (2012) 1250024; F.G. Alvarenga, M.J.S. Houndjo, A.V. Monwanou, Jean B.Chabi Orou, J. Mod. Phys. {\bf 4} (2013) 130-139; M.J.S. Houndjo, F.G. Alvarenga, Manuel E. Rodrigues, Deborah F. Jardim, Eur. Phys. J. Plus {\bf 129} (2014) 171. 
\bibitem{fRTf} F.G. Alvarenga, A. de la Cruz-Dombriz, M.J.S. Houndjo, M.E. Rodrigues, D. S\'aez-G\'omez, Phys. Rev. D {\bf 87} (2013) no.10, 103526; E.H. Baffou, A.V. Kpadonou, M.E. Rodrigues, M.J.S. Houndjo, J. Tossa, Astrophys.Space Sci. {\bf 356} (2015) no.1, 173-180; E.H. Baffou, M.J.S. Houndjo, M.E. Rodrigues, A.V. Kpadonou, J. Tossa,  Chin.J.Phys. {\bf 55} (2017) 467; E.H. Baffou, I.G. Salako, M.J.S. Houndjo, Int.J.Geom.Meth.Mod.Phys, {\bf 14} (2017) no.04, 1750051.
\bibitem{fTi} Kazuharu Bamba, Ratbay Myrzakulov, Shin'ichi Nojiri, Sergei D. Odintsov, Phys. Rev. D{\bf 85} (2012) 104036; Jaume Amor\'os, Jaume de Haro, Sergei D. Odintsov, Phys. Rev. D {\bf 87} (2013) 104037; Kazuharu Bamba, Shin'ichi Nojiri, Sergei D. Odintsov, Phys. Lett. B{\bf 731} (2014) 257-264; Kazuharu Bamba, Sergei D. Odintsov, Emmanuel N. Saridakis, Mod. Phys. Lett. A {\bf 32} (2017) no.21, 1750114. 
\bibitem{fTf} M. Hamani Daouda, Manuel E. Rodrigues, M.J.S. Houndjo, Eur. Phys. J. C {\bf 71} (2011) 1817; M. Hamani Daouda, Manuel E. Rodrigues, M.J.S. Houndjo, Eur. Phys. J. C {\bf 72} (2012) 1890; M.E. Rodrigues, M.J.S. Houndjo, D. Saez-Gomez, F. Rahaman, Phys. Rev. D {\bf 86} (2012) 104059; I.G. Salako, M.E. Rodrigues, A.V. Kpadonou, M.J.S. Houndjo, J. Tossa, JCAP {\bf 1311} (2013) 060
\bibitem{22debamba} J. L. Anderson and D. Finkelstein, Am. J. Phys. {\bf 39}, 901 (1971); W. Buchmuller and N. Dragon, Phys. Lett. B
{\bf 207}, 292 (1988); M. Henneaux and C. Teitelboim, Phys. Lett. B {\bf 222}, 195 (1989); W. G. Unruh, Phys. Rev. D
{\bf 40}, 1048 (1989); Y. J. Ng and H. van Dam, J. Math. Phys. {\bf 32}, 1337 (1991); D. R. Finkelstein, A. A. Galiautdinov
and J. E. Baugh, J. Math. Phys. {\bf 42}, 340 (2001); E. Alvarez, JHEP {\bf 0503}, 002 (2005); E. Alvarez,
D. Blas, J. Garriga and E. Verdaguer, Nucl. Phys. B {\bf 756}, 148 (2006); A. H. Abbassi and A. M. Abbassi, Class. Quant. Grav. {\bf 25}, 175018 (2008). 
\bibitem{int} G. F. R. Ellis, H. van Elst, J. Murugan and J. P. Uzan, Class.
Quant. Grav. {\bf 28}, 225007 (2011); P. Jain, Mod. Phys. Lett. A {\bf 27}, 1250201 (2012); N. K. Singh, Mod. Phys.
Lett. A {\bf 28}, 1350130 (2013); C. Barcel\'o, R. Carballo-Rubio and L. J. Garay, Phys. Rev. D {\bf 89}, 124019 (2014);
C. Gao, R. H. Brandenberger, Y. Cai and P. Chen, JCAP {\bf 1409}, 021 (2014); C. Barcel\'o, R. Carballo-Rubio
and L. J. Garay, arXiv:1406.7713 [gr-qc]; A. Padilla and I. D. Saltas, Eur. Phys. J. C {\bf 75}, 561 (2015); I. D. Saltas,
Phys. Rev. D {\bf 90}, 124052 (2014); J. Kluson, Phys. Rev. D {\bf 91}, 064058 (2015); A. Eichhorn, JHEP {\bf 1504}, 096 (2015);
E. Alvarez, S. Gonz\'alez-Mart\'ın, M. Herrero-Valea and C. P. Mart\'ın, JHEP {\bf 1508}, 078 (2015); A. Basak,
O. Fabre and S. Shankaranarayanan, arXiv:1511.01805 [gr-qc]; D. J. Burger, G. F. R. Ellis, J. Murugan and
A. Weltman, arXiv:1511.08517 [hep-th].
\bibitem{bamba} S.D. Odintsov, V.K. Oikonomou Astrophys. Space Sci. {\bf 361} (2016) no.7, 236; Kazuharu Bamba, Sergei D. Odintsov, Emmanuel N. Saridakis, Mod.Phys.Lett. A {\bf 32} (2017) no.21, 1750114; M.J.S. Houndjo, Eur.Phys.J. C77 (2017) no.9, 60; S. Nojiri, S.D. Odintsov, V.K. Oikonomou, Mod.Phys.Lett. A {\bf 31} (2016) no.30, 1650172; S. Nojiri, S.D. Odintsov, V.K. Oikonomou, Phys.Rev. D{\bf 93} (2016) no.8, 084050; S. Nojiri, S.D. Odintsov, V.K. Oikonomou, JCAP {\bf 1605} (2016) no.05, 046.
\bibitem{24debamba} I. Cho and N. K. Singh, Class. Quant. Grav. {\bf 32}, 135020
(2015).
\bibitem{25debamba} P. Jain, P. Karmakar, S. Mitra, S. Panda and
N. K. Singh, JCAP {\bf 1205}, 020 (2012); P. Jain, A. Jaiswal,
P. Karmakar, G. Kashyap and N. K. Singh, JCAP {\bf 1211},
003 (2012).
\bibitem{setareft} Xing Wu, Z.-H. Zhu, Phys. Lett. B {\bf 660}, 293–298 (2008).
\bibitem{oliver} M.J.S. Houndjo, O.F. Piattella,  {\bf 21}, (2012) 1250024. 
\bibitem{chinois}  M.R. Setare, F. Darabi, arXiv:1110.3962v1 [physics.gen-ph].



\end{thebibliography}
\end{document}